# *Velocity profile of granular flows down a heap described by dimensional analysis*

E. Martinez[a], O. Sotolongo-Costa[a,b], A. J. Batista-Leyva[a,c] and E. Altshuler[a]

a) Group of Complex Systems and Statistical Physics, Physics Faculty, University of Havana, 10400, Havana, Cuba
b) Universidad Autónoma del Estado de Morelos, Cuernavaca, Morelos, México.
c) Instituto Superior de Tecnologías y Ciencias Aplicadas (InSTEC), Salvador Allende, La Habana 10600, Cuba.

Gravity-driven thick granular flows are relevant to many industrial and geophysical processes. In particular, it is important to know and understand the particle velocity distributions as we get deeper into the flow from the free surface. In this paper, we use dimensional analysis as a tool to reproduce the velocity profile experimentally reported for granular flows down a confined heap for the so-called flowing layer and the creep layer underneath: the grains velocity first decrease linearly from the free surface, and then exponentially.

**See the published version of this paper at http://rcf.fisica.uh.cu/index.php/es/2015-07-15-15-25-47/volumen-32-numero-1-2015**

## 1. Introduction

Agriculture and food industries, minery, metallurgy, transportation, construction and pharmacology are examples of human activities where granular materials are basic ingredients [1]. On the other hand, the study of granular matter has shown to be a source of concepts and theoretical methods useful in many branches of research [2].

Granular flow is one of the most important mechanisms in granular dynamics, although it is not easy to describe. Three regimes can be defined: a) jammed (or static) state, where the grain´s inertia is not essential, b) "gaseous" state, where interaction through binary collisions dominates the dynamics and c) dense flow regime, which can be described through "hydrodynamic" equations.

This last regime is commonly present in our life: avalanches, landslides, sand dunes displacement and flows of grains through funnels, hoppers or silos are some examples [3]. For that reason many experimental and theoretical work have been devoted to understand it in detail $[4-6]$.

Here, we will concentrate on the description of dense flows down a heap (figure 1 a): the fluid layer of granular matter is moving on a heap of nearly static grains. Two vertical glass plates (separated by a gap $W$) confine the grains and the flow rate is regulated by an external hopper. The pile slope in a steady flow is self-adjusted as a result of the stress distribution inside the heap an the friction against the walls. *The velocity profile is composed by an upper linear part (near the free surface) followed by an exponential tail deeper inside the granular flow. The distance from the free surface to the boundary between the two regions is given by* $h \sim \left(Q/(d\sqrt{gd})\right)^{1/2}$, *where $d$ is the grain´s diameters, $g$ is gravity's acceleration and $Q$, the input flux of the granular material (given as area/time).*[6] .

The flow is typically assumed as dense which facilitates to model it using hydrodynamics-like equations. This characterization raises many difficulties because of the lack of detailed information on the microscopic behavior of the grains. Nonetheless, for the description and interpretation of the experimental results (in particular the velocity profiles) a simple but powerful method has been used: *the dimensional analysis* $[5,7]$. The key to apply it is to identify which are the essential parameters of a particular phenomenon. Magnitudes related amongst themselves (*e. g.* density, mass and volume) are rarely included at a time in the set of relevant parameters and the goal is obtain a set of *independents* magnitudes to describe the physics of the problem. The following step is to select a set of basic units and search for the possible adimensional numbers using the Buckingham´s Pi theorem, so as to obtain the functional dependencies among the magnitudes describing the granular medium $[8-11]$.

Previous works $[5,6,12]$ have built a local rheology and developed a "microscopic" model for grain-grain interactions using dimensional analysis and an empirical law for the effective friction ($\mu_{eff}$). The experiments have revealed that "microscopic" parameters such as grain friction, shape and effective restitution coefficient, are somehow included into $\mu_{eff}$ $[13]$. In $[12]$ the authors propose that the basics magnitudes controlling these flows are: grain size ($d$), normal stress ($P$), shear stress ($\tau$), shear rate ($\dot{\gamma}$) and effective density of the flow ($\rho$). The influence of the gravity and the flow's height were included in $P$, $\tau$ and $\dot{\gamma}$. From those parameters they obtained two adimensional numbers: $\tau/P$ and $I = \dot{\gamma}d/\sqrt{P/\rho}$. The first one is interpreted as the effective friction and the second is the ratio of two different timescales at the grain level: $1/\dot{\gamma}$ and $d\sqrt{\rho/P}$. In figure 2 appears a simplified sketch ilustrating the meaning of these timescales. The number $I$ represents the "competition" between the inertia and the confining effects on grains. Assuming an empirical expression for $\mu_{eff}(I)$, it was possible to find a Bagnold-like velocity profile for an inclined plane and the linear velocity profile in the shear plane and in the heap, but it was not possible to obtain the exponential tail in this last configuration [5,6,12].

## 2. Results and discussion

Our objective is to find the two velocity profiles in a heap confined between vertical plates under a unique description. For that, we consider a non-local rheology where an order parameter ($\psi \epsilon [0,1]$) is varied to determine the influence of the static grains on the fluid layer [14]. Granular flow is regularly thought to be visco-plastic [12], thus the stress tensor ($\sigma$) is modeled as the sum of three components [14]: a viscous term ($\eta\dot{\gamma}$), where $\eta$ is the viscosity, the hydrodynamic pressure ($P$) and a stress ($\tau_0$) corresponding to the quasi-static behavior. We can evaluate the influence of this last parameter through the expression ($\psi\tau_0$), that we use in the calculation of $\sigma = \eta\dot{\gamma} + P + \psi\tau_0$. When $\psi = 0$, the granular flow´s behavior resembles a viscous flow; if $0 < \psi \leq 1$ and $\eta\dot{\gamma} + P \sim \tau_0$, we have a viscoplastic flow, else ($\eta\dot{\gamma} + P \ll \tau_0$), the flow is in a cuasi-static regimen [13].

Finally, we assume (as other authors [5,14]) that the flow of grains can be described by the Navier-Stokes equation using the previous stress tensor. Thus, the parameters that appear in this equation are the main magnitudes controlling the properties of the flows. Moreover, we must include the size of the grain as the only granular lenghtscale in the dimensional analysis. The influence of other lengths related with the geometry is established either by a macroscopic parameter like $P$ or a "microscopic" one such as $\tau_0, V$(velocity of grains) or $\dot{\gamma}$. Likewise, these parameters determine the effect of the flow rate variation. As the flows mentioned are described in a steady state we do not take the time into account. In summary, the magnitudes employed in the analysis are: $d, g, \rho, V$ and $\sigma$. For different dependences of $\sigma$ with $\psi$, we can obtain the different shapes of the velocity profile.

CASE 1: "Visco-plastic" fluid ($0 < \psi \leq 1$ and $\eta\dot{\gamma} + P \sim \tau_0$).

In this flow we can observe two kinds of flowing layers: one, close to the free surface where the grains behave like a granular "gas" and another, below the previous layer, where the grains flows like a liquid, experiencing enduring contacts [5, 6, 12]. Our analysis concerns the latter.

Here we introduce the magnitude $\tau_0$ and we take $\eta$ as an independent magnitude in the analysis. The velocity not only depends on $P$, but also on the nature of the "microscopic" inter-grains interaction. Therefore $V$ appears as an independent parameter as well. Finally, we note that gravity influences both $P$ and the inter-grains friction. In conclusion, we use $d, \rho, \dot{\gamma}$, $P$, $\tau_0$, $\eta$, $V$ and $g$ as our set of basic magnitudes. In order to diminish the number of adimensional combinations we enlarged the number of basic units, taking into account that the length's dimension changes according to the direction, thus, they are $L_x, L_y$ or $L_z$ [11]. We will use the coordinate system represented in figure 1 b to denote the $x$, $y$ and $z$ directions, where the last one is perpendicular to the paper. We will employ only the component of the gravity´s accelerations in the $y$ direction, because it is an experimental fact that in the $x$ direction the velocity of the grains flowing in a heap is constant [15] as a result of the compensation of the effect of the component of $g$ in this axis and the dissipative shocks between grains.

Then, we have five basic units (the previous plus mass and time) and we can construct only three adimensional numbers: $\tau_0/P$, $\tau_0/\rho\, d\, g$ and $\eta\dot{\gamma}/\rho\, d\, g$. We draw the constancy of the shear rate ($\dot{\gamma} = \partial V/\partial y$) from the following expression:

$$\dot{\gamma} = C_a\, \frac{\rho\, d\, g}{\eta}\, F(\frac{\tau_0}{P}, \frac{\tau_0}{\rho\, d\, g}, \theta_c) \qquad (1)$$

Here, $C_a$ is an adimensional constant and the functional shape of $F$ is unknown for dimensional analysis. Besides, the magnitude $\theta_c$ (the angle of repose of the heap) must be included as another adimensional number that characterizes the interactions between grains.

As the height of the flow increases, the friction against the wall and pressure rises too; we can therefore keep the assumption about the independence of $\tau_0/P$ with height of the fluid layer, based on the fact that the influence of the wall is taken into account in $\tau_0$. *Unlike $\tau_0/P$, the rate $\tau_0/\rho dg$ does not include the competition between tangential and normal tensions, but only the effect of the inter-grain and grain-wall frictions. In reference [6] the authors conclude: "The velocity gradient inside the flowing layer is identical for three different materials (glass, steel and aluminium)". That means that the variation of the friction details does not produce changes: modifications in both $P$ and $\tau_0$ can occur in such a way that their ratio keeps constant independently from $y$. We can then conclude that $\mathbf{F}$ does not depend significantly on $\tau_0/\rho dg$ at different heights of the flow.* So, if we integrate (1) over the $y$ direction we obtain a linear velocity profile.

CASE 2: Cuasi-static regimen ($0 < \psi \leq 1$ and $\eta\dot{\gamma} + P \ll \tau_0$)

In this configuration the most important mechanism of interaction between grains is the frictional contact. This implies that the shear of the creep layer and pressure are less significant than the *grain-grain* and *grain-wall* frictions, both included in $\tau_0$. Again $g, d$ and $\rho$ are relevant to our scenario. Besides, we take into account the velocity of grains in this layer and, in order to establish a difference with the typical velocity $V$ associated to the fluid layer, we named the former as $V_{creep}$. Although we do not include the shear rate, it is a matter of fact that there is a velocity gradient along the height of this layer which we must include in the analysis as $\partial V_{creep}/\partial y$. This expression is mathematically identical to $\dot{\gamma}$, but the physical meaning of these magnitudes are different: $\partial V_{creep}/\partial y$ includes, not only the effect of the shear stress or the pressure, like $\dot{\gamma}$, but also includes the contact friction between the grains.

Given $g, d, \rho, \tau_0, V_{creep}$ and $\partial V_{creep}/\partial y$ and the appropriate number of basic units (five, as before) we can construct just one adimensional combination. In the Appendix **1** we show that only if we use the characteristic size of the grains in the direction of the $y$-component of the gravity (along the height of the layer) we can produce the observed exponential velocity profile [5,6,16]. Using a different direction of the grain's typical size entails incorrect functional dependences as shown in Appendix **1.**

The adimensional number we refer to is $(d/V_{creep})(\partial V_{creep}/\partial y)$. After integrating this expression we obtain the profile:

$$V_{creep} \sim e^{\widehat{C_a}\frac{y}{d}} \qquad (2)$$

where $\widehat{C_a}$ is an adimensional constant, that depends of $\theta_c$, and the zero of the $y$ axis is taken at the interface between the flowing and creep layers. In agreement with the experimental results, in (2) appears that the "attenuation length" of the velocity is of the order of $d$.

Our procedure is also corroborated by this fact: when the grain employed are not spherical but ellipsoidal, the semi-axis defined as the grain's "diameter" in order to adjust the exponential velocity profile, has to be parallel to the direction along the height of the layer, according to our selection of the $d$ dimension ($L_y$) [16].

As a final result, in the Appendix **2** we apply dimensional analysis to another geometries that had been extensively studied, and we obtain the same velocities profiles reported in the literature for these configurations [5,5,11].

## 3. Conclusions

With all these results we can do an interpretation of the transition between one velocity profile functional dependence to the other, which is an outcome of the variation of granular packing. The experimental validation can be done in the flow on a heap because two velocity profiles coexist. It is reported that the volume fraction is much higher in the creep layer (which is, besides, roughly constant here) than in the flowing layer (which decreases quickly) [5]. Thus, in the former layer are more significant the effects of the forces between grains, which is considered here through the values of the order parameter $\psi$.

Using dimensional analysis we have been able to reproduce the velocity profile experimentally reported for granular flows down a confined heap for the flowing layer and the creep layer underneath: the grains velocity first decrease linearly from the free surface, and then exponentially.

Our choice of parameters for the dimensional analysis of the flowing layer is based on the experimental fact that there the interactions between grains occurs *via* collisions and shear. In the case of the creep layer, we assume that the interactions are mainly frictional, which is expected from the higher, and constant, packing fraction in that region.

**References.**

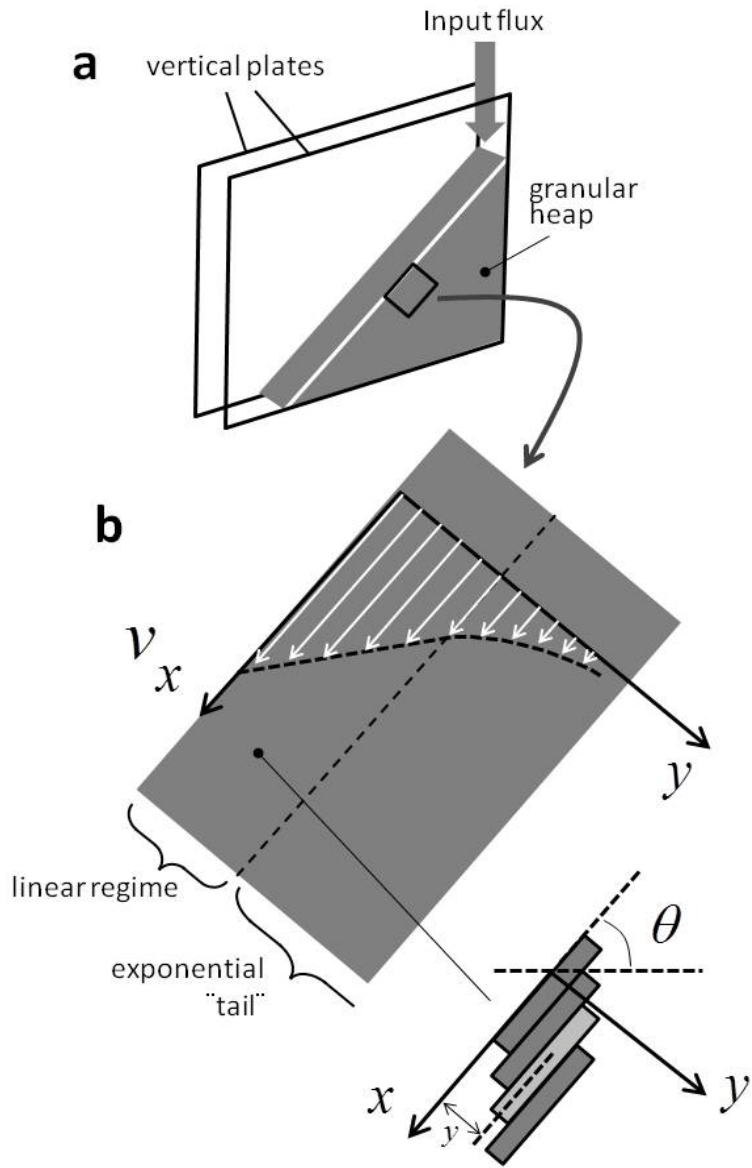

Figure1. Schematic representation of the velocity profile in a heap. a) A heap confined in Hele-Shaw cell. b) Detail of the velocity profile in a steady-state granular flow: the upper part is linear; deeper we can find an exponential tail. We can consider the granular flow as a many layer of grains slipping one above the other. The $z$-axis is perpendicular to the paper.

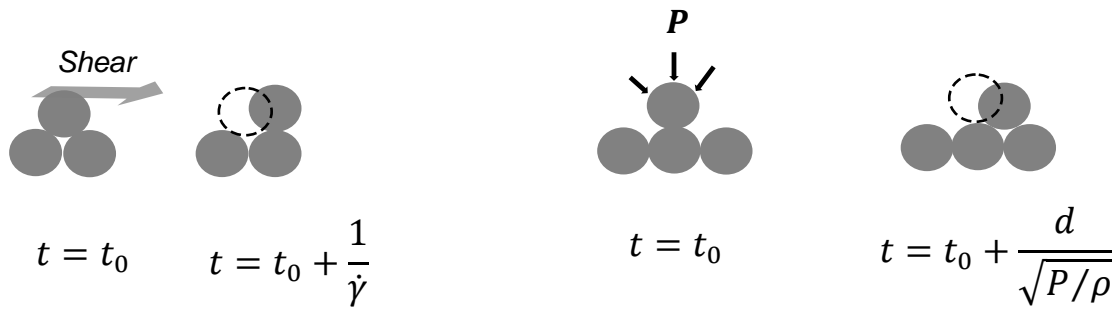

Figure 2. Simple diagramation of the physical meaning of two different timescales at the grain level. The interval $\tau_{shear} = 1/\dot\gamma$ represents the time needed by the grains to "climb" over a next particle because of the shear stress. Other interpretation is the macroscopic time needed for one layer to travel respect the other. In case of $\tau_{pressure} = d/\sqrt{P/\rho}$, represent the time needed by the grain to "fall" into a hole at a lower position due to the confining pressure.

# Appendix 1: Details of the dimensional analysis

## Case 1 (fluid layer in the heap)

Magnitudes: $V, \dot{\gamma}, \eta, \tau_0, g, d, \rho$.

Fundamentals dimensions: $L_x, L_y, L_z, T, M$

Dimensional matrix:

|    | V  | $\dot{\gamma}$ | $\eta$ | | | $\tau_0$ | | | g | d | | | $\rho$ |
|    |    |    | I  | II | III | I  | II | III |    | $I^1$ | $II^1$ | $III^1$ |    |
|----|----|----|----|----|-----|----|----|-----|----|----|----|----|----|
| Lx | 1  | 1  | -2 | -1 | -1  | -1 | 0  | 0   | 0  | 1  | 0  | 0  | -1 |
| Ly | 0  | -1 | 1  | 0  | 1   | 0  | -1 | 0   | 1  | 0  | 1  | 0  | -1 |
| Lz | 0  | 0  | 0  | 0  | -1  | 0  | 0  | -1  | 0  | 0  | 0  | 1  | -1 |
| T  | -1 | -1 | -1 | -1 | -1  | -2 | -2 | -2  | -2 | 0  | 0  | 0  | 0  |
| M  | 0  | 0  | 1  | 1  | 1   | 1  | 1  | 1   | 0  | 0  | 0  | 0  | 1  |

*NOTE*: We get the dimensions of the viscosity through the ratio of the dimensional formulas for the shear stress and the shear rate. That means that we must consider only the related combinations in the dimensional matrix.

Rank: **4**, for the combinations III-$I^1$ y III-$II^1$; **5** for the other ones. Just in the combination I-$III^1$ the magnitudes $V$ $and$ $\dot{\gamma}$ appears in only one of the two dimensional number that could be formed. For the other combinations these variables appear in both dimensional numbers and we can´t determine completely the functional dependence of the velocity profile.

Adimensional number, without $V$ $and$ $\dot{\gamma}$, for the combination I-$III^1$: $\dfrac{\tau_0}{\rho\, d\, g}$

Adimensional number, with $V$ $and$ $\dot{\gamma}$, for the combination I-$III^1$: $\dfrac{\eta\, \dot{\gamma}}{\rho\, d\, g}$

# Case 2 (creep layer in the heap)

Magnitudes: $V_{creep}, \frac{\partial V_{creep}}{\partial y}, \tau_0, g, d, \rho.$

Fundamental dimensions: $L_x, L_y, L_z, T, M$

Dimensional matrix:

|     | $V_{creep}$ | $\frac{\partial V_{creep}}{\partial y}$ | $\tau_0$ I | $\tau_0$ II | $\tau_0$ III | $g$ | $d$ I$^1$ | $d$ II$^1$ | $d$ III$^1$ | $\rho$ |
|-----|-------------|------------------------------------------|------------|-------------|--------------|-----|-----------|------------|-------------|--------|
| Lx  | 1           | 1                                        | -1         | 0           | 0            | 0   | 1         | 0          | 0           | -1     |
| Ly  | 0           | -1                                       | 0          | -1          | 0            | 1   | 0         | 1          | 0           | -1     |
| Lz  | 0           | 0                                        | 0          | 0           | -1           | 0   | 0         | 0          | 1           | -1     |
| T   | -1          | -1                                       | -2         | -2          | -2           | -2  | 0         | 0          | 0           | 0      |
| M   | 0           | 0                                        | 1          | 1           | 1            | 0   | 0         | 0          | 0           | 1      |

Rank: **4,** for the combinations III-I$^1$ y III-II$^1$; only in the first one is possible construct two adimensional number where in only one appears $V$ and $\dot{\gamma}$. **5** is the rank for the others combinations.

Adimensional number for the combination I-I$^1$, II-I$^1$: $\dfrac{d^2 g \frac{\partial V_{creep}}{\partial y}}{V_{creep}^3}$

Adimensional number for the combination I-II$^1$, II-II$^1$: $\dfrac{d \frac{\partial V_{creep}}{\partial y}}{V_{creep}}$

Adimensional number for the combination I-III$^1$: $\dfrac{d \rho g}{\tau_0}$

Adimensional number for the combination II-III$^1$: $\dfrac{\rho^2 d^2 \frac{\partial V_{creep}}{\partial y} V_{creep} g}{\tau_0^2}$

Adimensional number for the combination III-III$^1$: $\dfrac{\rho^2 g V_{creep}^3}{\frac{\partial V_{creep}}{\partial y} \tau_0^2}$

Adimensional number, without $V creep$ and $\frac{\partial V_{creep}}{\partial y}$, for the combination III-I$^1$: $\dfrac{d \rho g}{\tau_0}$

Adimensional number, with $V creep$ and $\frac{\partial V_{creep}}{\partial y}$, for the combination III-I$^1$: $\dfrac{d^2 g \frac{\partial V_{creep}}{\partial y}}{V_{creep}^3}$

# Appendix 2: Other flow geometries

There are other two relevant configurations (figure 3) where a simple shear flow is produced and its rheological properties can be measured [6,12] .

The most simple of these geometries is the plane shear (figure 3 a): the shear is produced by the motion of the upper wall. For reasons of simplicity the influence of the gravity ($g$) is neglected and thereby the stress distribution is uniform in all the extension of the fluid [6,12]. The velocity of the grains increases linearly in proportion to the height of the fluid layer.

Another well studied configuration is the flow on inclined plane (figure 3 b and c), which is frequent in both geophysical and industrial circumstances. The set-up is a rough and rigid plane at an angle (*θ)* over the horizontal and the flow rate is controlled through the opening of a gate fixed at the top of the plane. The experiments are conducted in such a way that the motion of the grains is stationary: in the steady-state, a uniform flow of height ($h$) is established and gravity acts as the driving force. The density ($\rho$) is assumed constant and the stress distributions are $\sigma_{tan} = \rho g h \sin(\theta)$ and $\sigma_{nor} = \rho g h \cos(\theta)$: the former characterizes the tangential stress and the latter, the normal stress. The velocity profile in the flow for big $h$ is a Bagnold-like profile ($V \sim h^{3/2}$) and for small $h$ the shape of the profile is linear.

In all cases the flow is also regarded as dense, therefore the description is regularly conducted from a hydrodynamic perspective, besides we follow the same method presented in the paper to describe the velocity profile on a heap.

CASE 1A: "Viscous" fluid ($\psi = 0$).

*Geometry*: Plane shear.

Gravity is not included and the information about the velocity of grains and the viscosity are contained in $P$ and $\dot{\gamma}$ [5,6,17]. This means that the magnitudes involve are $d, \rho, \dot{\gamma}$ and $P$. We take as basic units: length ($L$), mass ($M$), and time ($T$). This makes possible to find the number $I$ and obtain the linear velocity profile experimentally reported [5,6,12].

*Geometry*: Inclined plane (big $h$)

The only difference with the previous geometry is in $P$. Here its value is not constant as in the plane shear, though it still contains the information about the height of the flow (implicitly, the velocity) and gravity ($P \sim \rho g h$). Once again, the magnitudes are $d, \rho, \dot{\gamma}$ and $P$, the basic units are the same as before, so we can obtain the number $I$ and the Bagnold-like profile as has been reported before [6,11]. The incline angle is an adimensional magnitude on which the number $I$ relies, but this functional relation can't be established by dimensional analisis.

CASE 2A: "Visco-plastic" fluid ($0 < \psi \leq 1$ and $\eta \dot{\gamma} + P \sim \tau_0$).

*Geometry*: Incline plane (small $h$)

The dependence here is the same as that obtained for the fluid region in the heap: a linear profile. We can interpret this result as follows: as the height of the flow is small, the action of the rough bottom is significative and acts as a source of frictional forces that affects at the most part of the flowing grains. For this reason appears explicitly the magnitude $\tau_0$

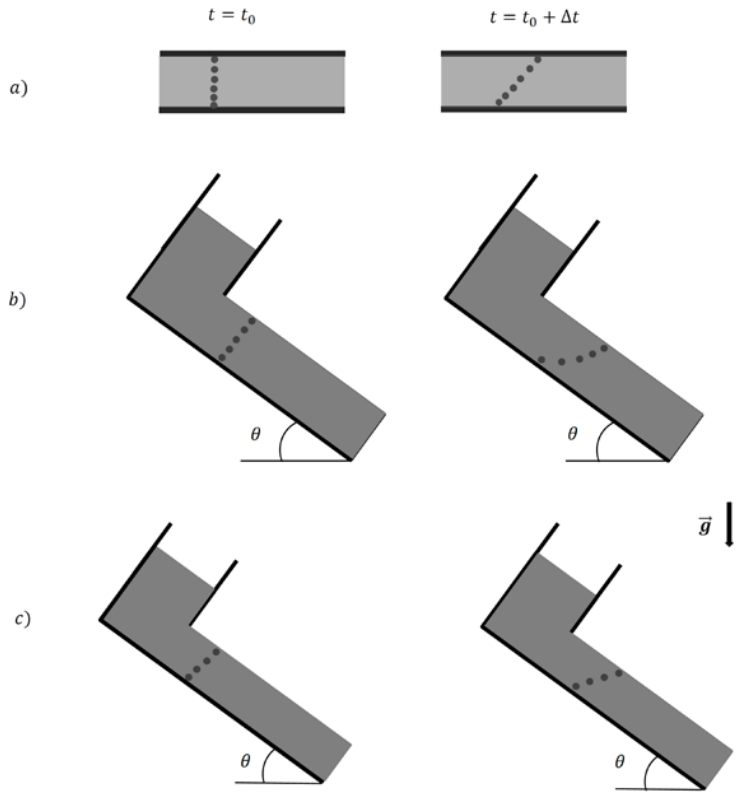

Figure 3. Another two common configurations to characterize granular flows. a) Plane shear, b) incline plane with a big height of the fluid layer and c) incline plane with a thin fluid layer. The dark spots are grains represented in two distinct time instants.